\documentclass{article}
\usepackage{graphicx}

\usepackage[final, nonatbib]{neurips_2023_ml4ps}
\usepackage[numbers]{natbib}

\usepackage[utf8]{inputenc} 
\usepackage[T1]{fontenc}    
\usepackage{hyperref}       
\usepackage{url}            
\usepackage{booktabs}       
\usepackage{amsfonts}       
\usepackage{nicefrac}       
\usepackage{microtype}      
\usepackage{xcolor}         
\usepackage{amsmath}

\title{Self-Driving Telescopes: Autonomous Scheduling of Astronomical Observation Campaigns with Offline Reinforcement Learning}

\author{
  Franco Terranova\\
  Department of Information Engineering\\
  University of Pisa\\
  Pisa, IT 56126 \\
  \texttt{terranovafr@icloud.com} \\
  \And
  M. Voetberg \\
  Computational Science and AI Directorate\\
  Fermi National Accelerator Laboratory\\
  Batavia, IL 60510 \\
  \texttt{maggiev@fnal.gov} \\
  \And
  Brian Nord \\
  Fermi National Accelerator Laboratory; \\Kavli Institute for Cosmological Physics \& \\ Department of Astronomy and Astrophysics,\\ University of Chicago;\\ 
  \texttt{nord@fnal.gov} \\
  \And
  Amanda Pagul \\
  Space Telescope Science Institute\\
  Baltimore, MD 21218\\
  \texttt{apagul@stsci.edu} \\
}

\begin{document}

\maketitle
\begin{abstract}
Modern astronomical experiments are designed to achieve multiple scientific goals, from studies of galaxy evolution to cosmic acceleration. 
These goals require data of many different classes of night-sky objects, each of which has a particular set of observational needs. 
These observational needs are typically in strong competition with one another.
This poses a challenging multi-objective optimization problem that remains unsolved.
The effectiveness of Reinforcement Learning (RL) as a valuable paradigm for training autonomous systems has been well-demonstrated, and it may provide the basis for self-driving telescopes capable of optimizing the scheduling for astronomy campaigns.
Simulated datasets containing examples of interactions between a telescope and a discrete set of sky locations on the celestial sphere can be used to train an RL model to sequentially gather data from these several locations to maximize a cumulative reward as a measure of the quality of the data gathered.
We use simulated data to test and compare multiple implementations of a Deep Q-Network (DQN) for the task of optimizing the schedule of observations from the Stone Edge Observatory (SEO). 
We combine multiple improvements on the DQN and adjustments to the dataset, showing that DQNs can achieve an average reward of 87\%±6\% of the maximum achievable reward in each state on the test set.
This is the first comparison of offline RL algorithms for a particular astronomical challenge and the first open-source framework for performing such a comparison and assessment task.
\end{abstract}

\section{Introduction}
Scheduling observations for telescopes is a challenging task, known to be an NP-hard (Non-deterministic polynomial-time hard) problem. 
This level of complexity has also been highlighted in other studies \cite{hinze2016telescope}.
Optimizing telescope scheduling can significantly enhance the efficiency of the planning process and the effectiveness of surveys, which may be otherwise sub-optimal in the case of manual optimization due to the complexity of the problem.
We have developed and implemented a framework based on the PTAN (PyTorch AgentNet) library \cite{ptan} of RL tools and utilities to compare various RL algorithms to achieve this goal.
This framework can be used on any offline dataset -- i.e.,  data typically generated by an expert, a simulation, or a historical record of observations.
The offline dataset must contain observation variables, actions taken in response to those observations, and examples of the resulting data quality after these actions. 
The dataset is scanned and used by our framework to achieve an optimized observing schedule.
This study focuses on the results obtained from an offline simulation dataset based at the Stone Edge Observatory (SEO) \protect\hyperlink{seo-website-footnote}{\textsuperscript{1}}.
When dealing with an offline dataset, which has a limited amount of data, value-based methods tend to perform better than other methods since they are sample-efficient and well-suited for off-policy learning. \cite{sutton2018reinforcement} 
This consideration has been confirmed with our specific dataset, on which they overcome other classes of methods, and this is the reason Deep Q-Networks (DQNs) \cite{mnih2013playing} are the main targets of this work.
\begin{figure}[htbp]
  \centering
  \includegraphics[width=0.50\textwidth]{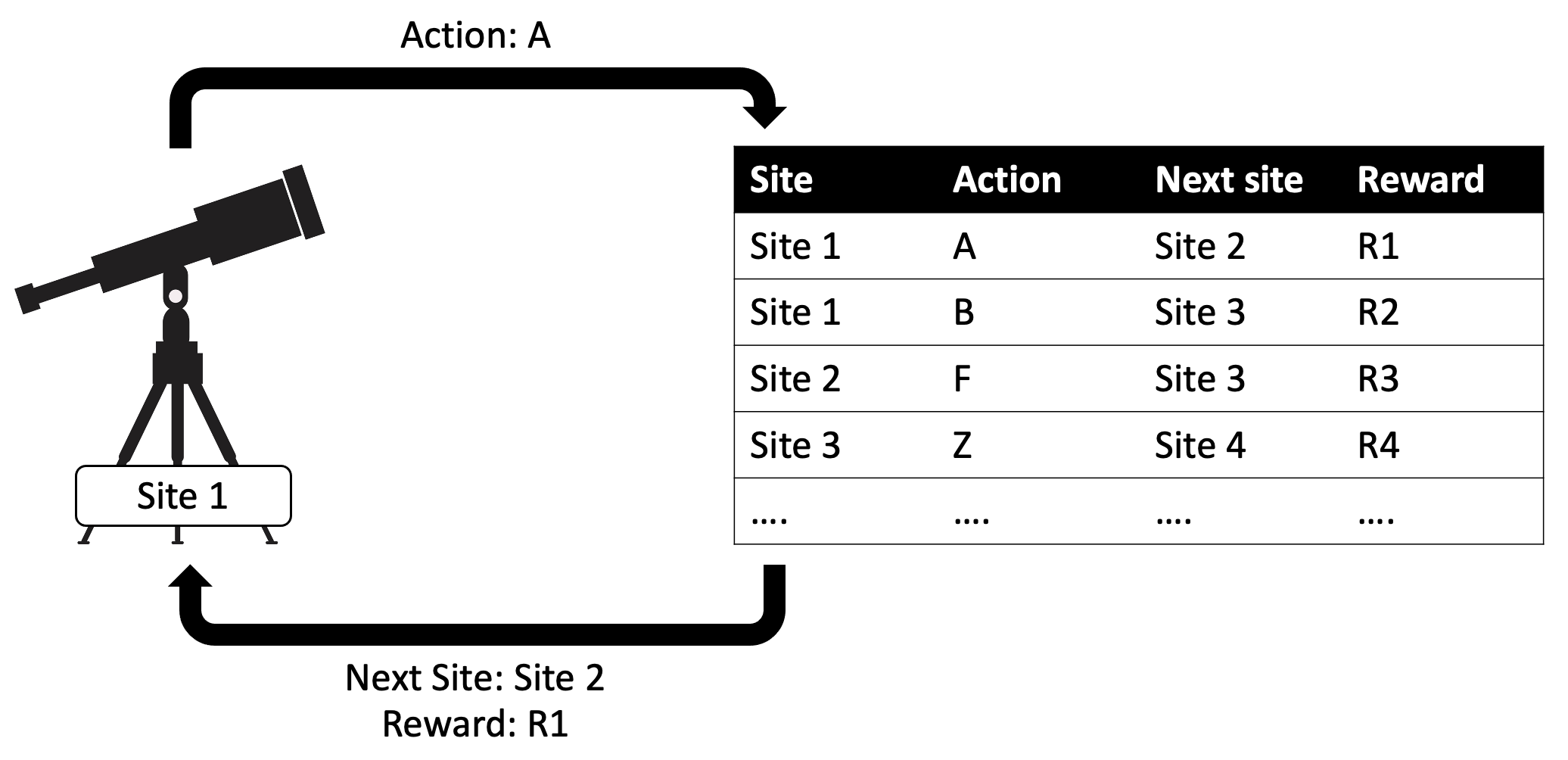}
  \caption{The RL paradigm applied to our task considers the telescope agent interacting with the environment represented by an offline dataset. The RL agent gathers the state, and executes an action, and the environment will respond with a reward and a new state.}
  \label{fig:RLParadigm}
\end{figure}
\hypertarget{seo-website-footnote}{}
\footnotetext[1]{SEO website: \url{https://sites.google.com/a/starsatyerkes.net/yerkesprojects/tools/telescope-upgrades/stone-edge-observatory?pli=1}}

\subsection{Limitations}
The analysis performed in this study can be biased by the characteristics of the specific dataset used in terms of generalization since it represents a specific simulation survey of a specific observatory. 
However, the framework we developed for this study can be used to compare multiple RL algorithms on any offline simulation dataset and can be used in the future to study the capacity for generalization of algorithms regardless of the offline dataset used.
Our simulated dataset contains only positions of sky objects, but not factors such as weather or climate conditions. 
Furthermore, when addressing multi-objective problems, optimizing a single data quality metric, as will be performed in this study, may fail to fully represent the problem, even if it considers a general metric for data quality. A more sophisticated reward function should be needed to achieve a specific objective.

\subsection{Related work}
In the field of self-driving telescopes and schedule optimization, most of the research has been primarily approached with heuristic methods, as demonstrated in prior work \cite{minton1990solving} \cite{naghib2016feature}.
In the study by Alba et al. \cite{alba2019sky}, it was observed that tabular approaches had their limitations in this context. 
On the other hand, Jia et al. \cite{jia2023observation} primarily focused on the challenges and solutions related to distributed telescope arrays.
More recently, automated approaches have been tested for large cosmic surveys: e.g., RL algorithms for the Rubin Observatory's Legacy Survey of Space and Time (LSST) \cite[e.g.,][]{naghib_framework_2019} and graph neural networks for spectroscopic surveys \cite[e.g.,][]{cranmer_unsupervised_2021}. 
All of these classes of techniques require an environment, in the real world or otherwise, to be trained. 

\section{Methodology}
We describe the methodology we built to address the application of value-based RL techniques on the benchmark SEO offline dataset, which contains a set of discrete locations in the sky. Consequently, it requires a discrete action space.
Datasets, checkpoints, and the code used are all available in Zenodo \cite{data}. 
For each algorithm and experiment, we make available the tensorboard logs and hyper-parameters. 
Hyper-parameters related to the problem and algorithms were selected based on discussion with domain experts and after manual comparison.

\subsection{Offline dataset}
This analysis uses an offline dataset that contains examples of interactions between the SEO telescope and the sky, which are generated through a simulation tool.
The offline dataset consists of 309,599 observations, representing a collection of state-action-reward-next state samples.
The data representing the observation space (i.e. state space) consists of several telemetry and positioning factors related to the telescope, the sun, and the moon.
The action space consists of Right Ascension and Declination coordinates.
In particular, the discrete set of unique actions available in the dataset corresponds to the sets of locations available for optimization.
The effective exposure time ($t_{effective}$)\cite{teff} is the reward variable for the RL agent, and it represents the depth (i.e.,  quality) of the images collected (Equation \ref{eq:teff}).
This variable indicates the level of brightness required for detecting the faintest point source at a given reference signal-to-noise ratio.
\begin{equation}
    \label{eq:teff}
    T_{\text{eff}} \equiv \eta^2 \left(\frac{0.9''}{\text{FWHM}}\right) \left(\frac{b_{\text{dark}}}{b}\right) *\tau_{exposure}
\end{equation}
The equation considers the key factors affecting exposure quality, including seeing (\textit{FWHM}, blurring caused by atmospheric and instrumental factors), atmospheric transmission (\(\eta\), the fraction of light reaching the instrument), and the sky brightness (\textit{b}), including a component related to low background light levels  \((b_{\text{dark}}\)). The component $\tau_{exposure}$ is the exposure time of a given observation. 
The selection of this reward model aims to drive the telescope to prioritize the optimization of the gathered data's quality.

\subsection{Action selection}
\label{chap:actions}

The dataset only contains rewards for a subset of all unique actions in different states.
Therefore, the agent might suggest telescope coordinates for which the reward is not available in the dataset in a certain state. 
Instead of dedicating an output neuron of the neural network for each unique action, which may be chosen in some states in which it will not be available, the dataset's right ascension and declination ranges have been evenly divided into 19 bins, and an output neuron is dedicated to each pair of bins.
The training buffer hence selects the action (sky location) closest available -- by angular distance -- to the chosen action.
This number of bins has been properly computed based on the dataset, in order to ensure that there's at least one output neuron mapped to each action, even in the worst-case scenario where all actions are available. 
This prevents the need to modify the network structure if the dataset expands.

\subsection{Evaluation metric: the average effectiveness distribution}
During testing, the reward distribution for each state is normalized to a range of [0, 1] to enable fair comparisons: different states have rewards with different scales. 
We refer to the average reward in an episode as "average effectiveness," and we calculated an average effectiveness distribution by conducting numerous episodes in both the full environment (1000 episodes) and the test set (500 episodes) to compare agents, ensuring result stability for agent comparisons. 
This distribution is generated using the combination of checkpoints that yielded the best average reward on the training and validation sets.

\subsection{DQN Network Architecture}
The number of neurons in the first layer of the DQN reflects the number of variables in the observation space (26 neurons). 
The architecture has three hidden layers (with 64, 128, and 256 neurons, respectively). The LeakyReLU activation function has been used with the amount of leak \(\alpha\) set to 0.01.
The number of bins was calculated as described in Section \ref{chap:actions}, resulting in 19 bins for the two action variables, hence a 361-dimensional (19x19) output space representing all possible pairs of bins. 

\section{Results}

\subsection{Hyper-parameters and normalization}
The neural network was trained for 300,000 iterations on three parallel training environments. 
The selections of hyper-parameters were the result of manual optimization and domain knowledge.
The replay buffer's size was set to 5,000 transitions, with a target network's synchronization period of 2,000 iterations.
For the exploration strategy, the Basic DQN includes an \(\epsilon\)-greedy strategy with a linear decay in 100,000 iterations from 1 to 0.05.
The holdout method is used with test and validation ratios set to 20\%. 
The combination of z-score normalization for the observation space variables and the intra-state reward distribution pushed the average effectiveness distribution mean value from 36\%±24\% to a value of 77\%±12\% on the test set for Basic DQN, demonstrating to be the most effective strategy to improve performances.  
Several hyper-parameters have also been varied, proving their impact on the performance of the network. 
In particular, the discounting factor (\(\gamma\)), set at 0.8, the learning rate at 0.001, and a batch size of 128 emerged as the most influential hyperparameters. 
It is crucial to highlight that the performance of a baseline random agent, which selects a random site within each state, yields an average effectiveness distribution with a mean value of 39\%±12\%.

\subsection{Improvements in literature}
\begin{figure}[htbp]
  \centering 
  \begin{minipage}[t]{0.42\textwidth}
    \includegraphics[width=\linewidth]{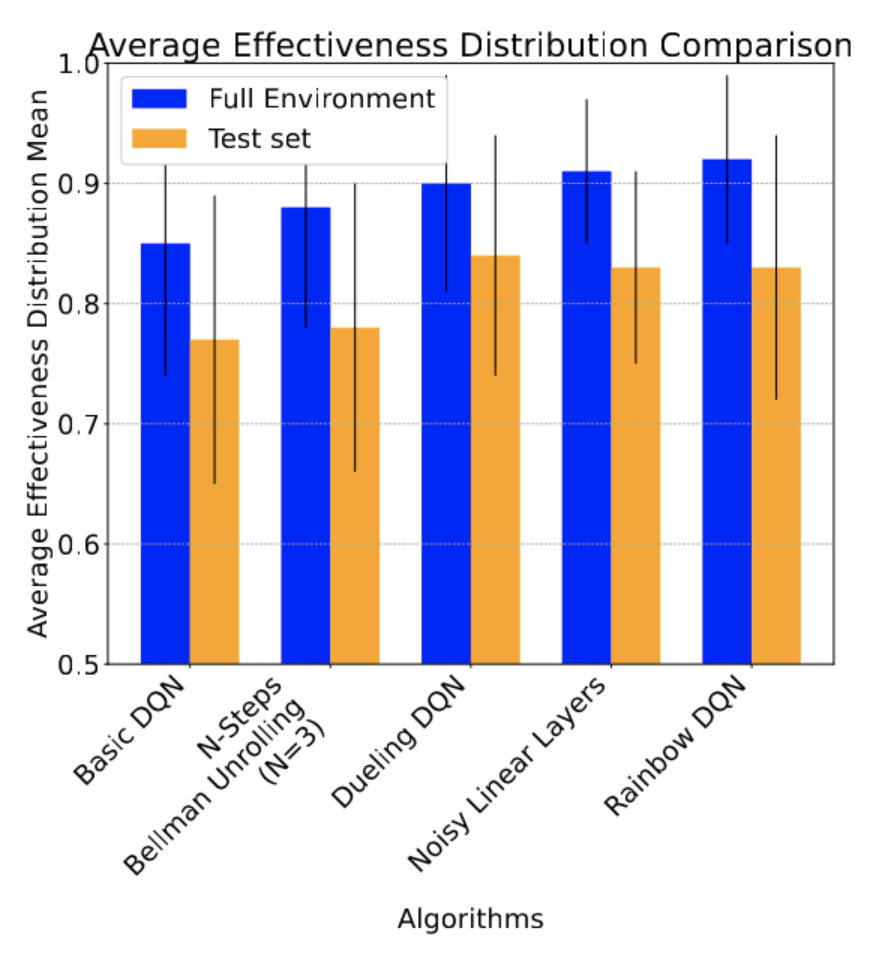}
    \caption{Comparison among improvements to the DQN method in terms of average effectiveness distribution.}
    \label{fig:improvements_dqn}
  \end{minipage}
  \hfill
  \begin{minipage}[t]{0.54\textwidth}
    \includegraphics[width=\linewidth]{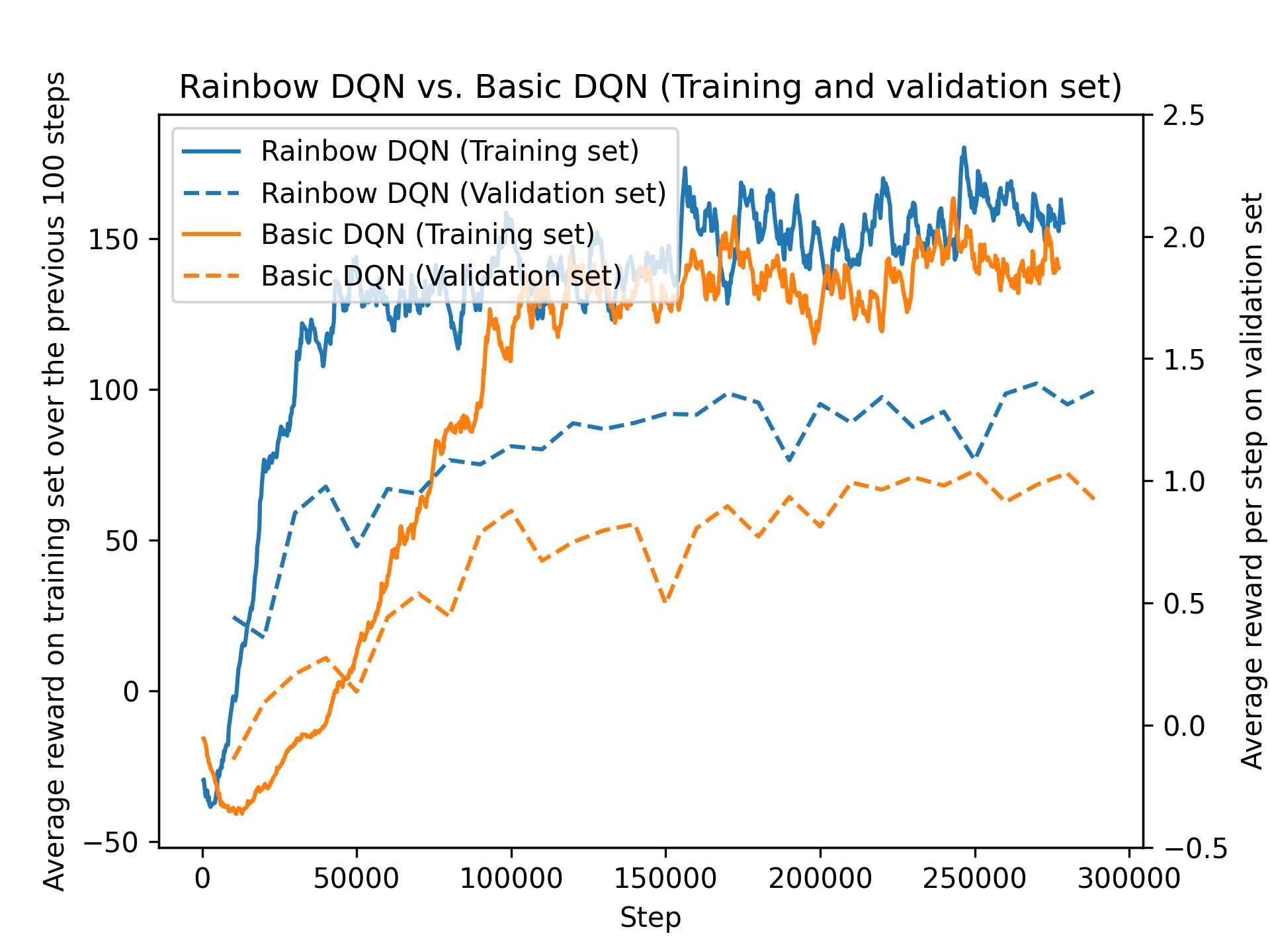}
    \caption{Average reward trend on the training set during the last 100 iterations and validation average reward per step during training: comparison of the Rainbow DQN and the Basic DQN.}
    \label{fig:avgreward_rainbow}
  \end{minipage}
\end{figure}

Further advanced modifications in the literature, including Dueling DQN \cite{wang2015dueling}, N-steps Bellman unrolling \cite{sutton1998learning}, and noisy layers exploration strategy \cite{fortunato2017noisy}, have been compared to highlight the improvements in the performance of the DQN model (Figure \ref{fig:improvements_dqn}).
These enhancements have been combined, analogously to the Rainbow DQN approach \cite{hessel2018combining}.
Figure \ref{fig:avgreward_rainbow} summarizes the comparison between the combination of the DQN improvements, or "Rainbow" DQN, and the original or "Basic" DQN, showing an improvement in general performance, convergence, and generalization capabilities during training.

\subsection{Evaluation}
\begin{figure}[htbp]
  \centering
  \begin{minipage}[t]{0.45\textwidth}
    \includegraphics[width=\linewidth]{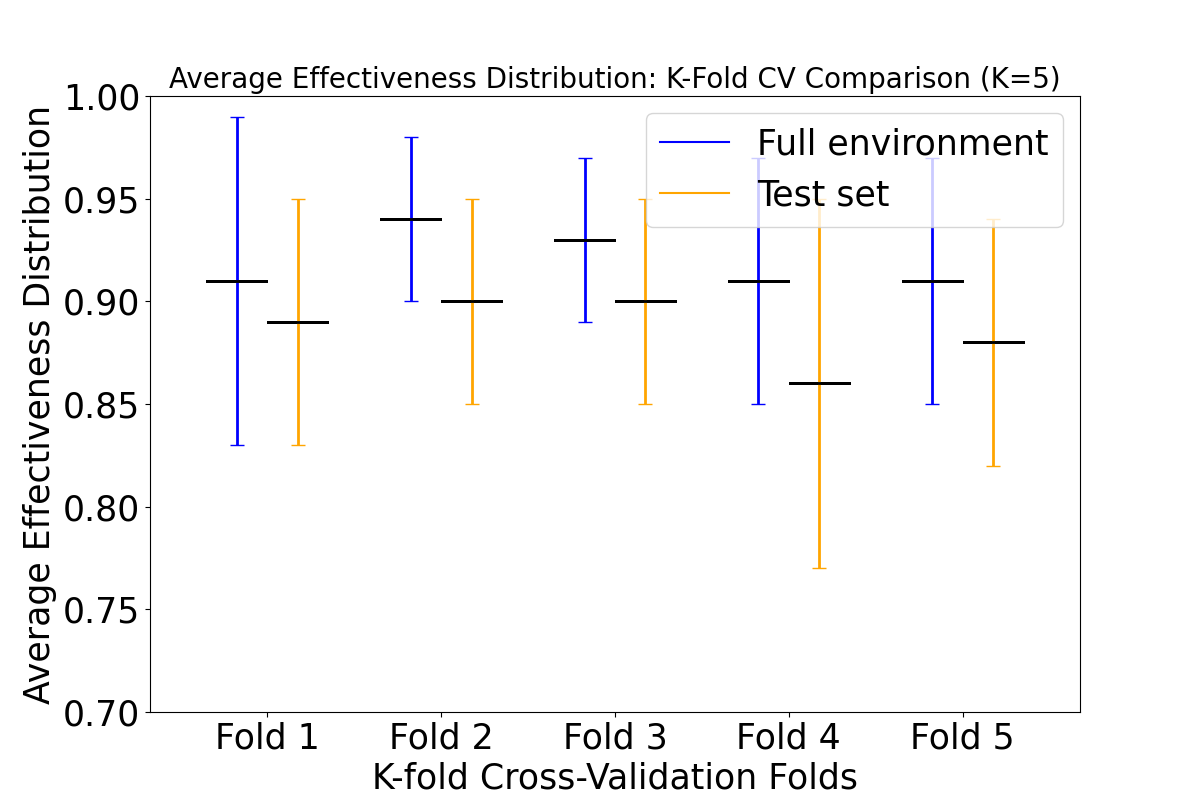}
    \caption{Rainbow DQN K-Fold CV average effectiveness distribution: comparison among folds results.}
    \label{fig:avg_effectiveness_kfold}
  \end{minipage}
  \hfill
   \begin{minipage}[t]{0.5\textwidth}
    \includegraphics[width=\linewidth]{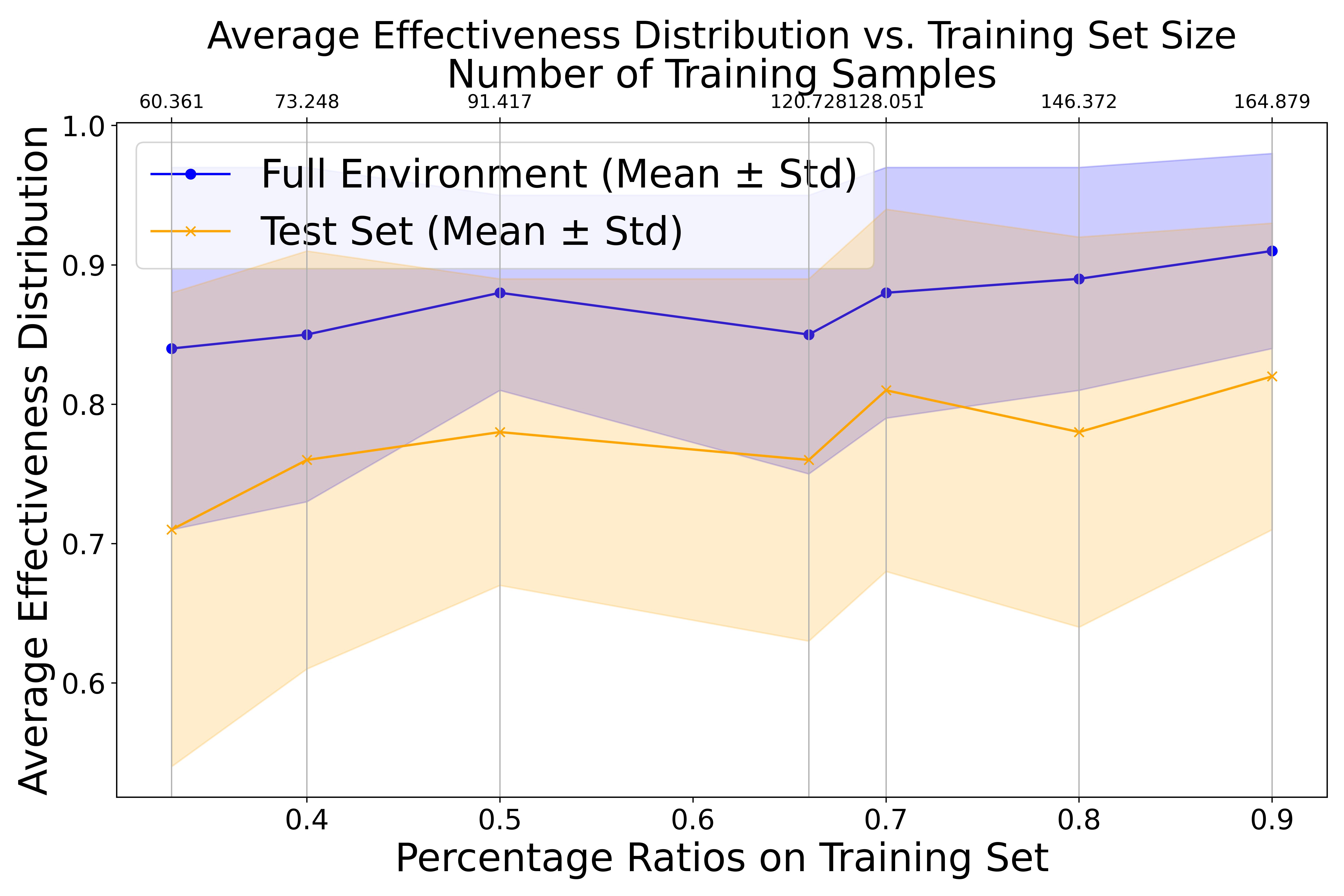}
    \caption{Average effectiveness distribution comparison: reduction of the percentage ratios of the training set.}
    \label{fig:samples_modification}
  \end{minipage}
\end{figure}

The Rainbow DQN has been assessed with a K-Fold Cross Validation (K-Fold CV) using K = 5, highlighting how the performances are pretty similar regardless of the portion of the dataset used for training, validating, or testing (Figure \ref{fig:avg_effectiveness_kfold}).
Moreover, the robustness of the model has been addressed by decreasing the quantity of samples in the training set. 
The resulting models trained with a reduced sample size have been compared on the same test set (Figure \ref{fig:samples_modification}), showcasing the model's capacity to learn effectively even with a lower number of samples, important for the particular context of learning from an offline dataset.

\section{Conclusions and Impact Statement}
Our results suggest that RL algorithms can be used to optimize the sequential schedule of a telescope survey. 
A proper framework has been developed for the purpose of wrapping offline datasets with the goal of using them as environments.
Value-based methods, and in particular DQNs, have shown remarkable success in tackling schedule optimization based on an offline simulation dataset located at the SEO. 
The combination of normalization techniques, hyperparameters exploration, and extensions to the DQN method have pushed the performances of the agent way further.
This research is expected to yield substantial advancements in the realm of enhancing telescope observations using RL. 
Even though this work primarily focuses on the domain of astronomical observations, the methodology discussed could be adapted to any problem related to schedule optimization.

\begin{ack}
This work was produced by Fermi Research Alliance, LLC under Contract No. DE-AC02-07CH11359 with the U.S. Department of Energy, Office of Science, Office of High Energy Physics. Publisher acknowledges the U.S. Government license to provide public access under the DOE Public Access Plan DOE Public Access Plan. 

We acknowledge the Deep Skies Lab as a community of multi-domain experts and collaborators who’ve facilitated an environment of open discussion, idea-generation, and collaboration. This community was important for the development of this project. 

We express our gratitude to Aleksandra \'Ciprijanovi\'c for her assistance in editing this manuscript and to Eric Neilsen Jr. for helping us with his expertise in delving into the problem's domain.

\end{ack}

\medskip

\nocite{*}
\bibliographystyle{ieeetr}
\bibliography{bibliography}

\end{document}